\documentclass[prd,showpacs,preprintnumbers]{revtex4} 

\usepackage{graphicx}
\usepackage{dcolumn}
\usepackage{bm}
\usepackage{latexsym}
\usepackage{amsfonts}
\usepackage{amssymb}
\usepackage{amsmath}

\begin{document}


\title{Modified Gravitational Equations on Braneworld with Lorentz Invariant Violation}
\author{Arianto$^{(1,2,3)}$}
\email{arianto@upi.edu}
\author{Freddy P. Zen$^{(1,2)}$}
\email{fpzen@fi.itb.ac.id}
\author{Bobby E. Gunara$^{(1,2)}$}
\email{bobby@fi.itb.ac.id}
\affiliation{$^{(1)}$Theoretical Physics Lab., THEPI Devision, \\
and \\
$^{(2)}$Indonesia Center for Theoretical and Mathematical Physics (ICTMP)\\
Faculty of Mathematics and Natural Sciences,\\
 Institut Teknologi Bandung,\\
Jl. Ganesha 10 Bandung 40132, INDONESIA. \\
$^{(3)}$Department of Physics, Udayana University\\
Jl. Kampus Bukit Jimbaran Kuta-Bali 80361, INDONESIA.\\
}

\begin{abstract}
The modified gravitational equations to describe a
four-dimensional braneworld in the case with the Lorentz invariant
violation in a bulk spacetime is presented. It contains a trace
part of the brane energy-momentum tensor and the coefficients of
all terms describe the Lorentz violation effects from the bulk
spacetime. As an application, we apply this formalism to study
cosmology. In respect to standard effective Friedmann equations on
the brane, Lorentz invariance violation in the bulk causes a
modification of this equations that can lead to significant
physical consequences. In particular, the effective Friedmann
equation on the brane explicitly depends on the equation of state
of the brane matter and the Lorentz violating parameters. We show
that the components of five-dimensional Weyl curvature are related
to the matter on brane even at low energies. We also find that the
constraints on the theory parameters are depend on the equation of
state of the energy components of the brane matter. Finally, the
stability of the model depend on the specific choices of initial
conditions and the parameters $\beta_i$.
\end{abstract}

\pacs{ 98.80.Cq, 98.80.Hw}
\maketitle

\section{Introduction}
There has been a growing appreciation of the importance of the
violations of Lorentz invariance recently. The intriguing
possibility of the Lorentz violation is that an unknown physics at
high-energy scales could lead to a spontaneous breaking of Lorentz
invariance by giving an expectation value to certain non Standard
Model fields that carry Lorentz indices, such as vectors, tensors,
and gradients of scalar fields~\cite{Kostelecky:1988zi}. A
relativistic theory of gravity where gravity is mediated by a
tensor, a vector, and a scalar field, thus called TeVeS
gravitational theory~\cite{Bekenstein:2004ne}, provides modified
Newtonian dynamics (MOND) and Newtonian limits in the weak field
nonrelativistic limit. TeVeS could also explain the large-scale
structure formation of the Universe without recurring to cold dark
matter \cite{Skordis:2005xk}, which is composed of very massive
slowly moving and weakly interacting particles. On the other hand,
the Einstein--Aether theory \cite{Jacobson:2000xp} is a
vector-tensor theory, and TeVeS can be written as a vector-tensor
theory which is the extension of the Einstein--Aether
theory~\cite{Zlosnik:2006sb}. In the case of generalized
Einstein--Aether theory~\cite{Zlosnik:2006zu}, the effect of a
general class of such theories on the solar system has been
considered in Ref.~\cite{Bonvin:2007ap}. On small scales the
Einstein-Aether vector field will in general lead to a
renormalization of the local Newton
Constant~\cite{Carroll:2004ai}. Moreover, as has been shown by
authors in Ref.~\cite{Li:2007vz}, the Einstein--Aether theory may
lead to significant modifications of the power spectrum of tensor
perturbation. The existence of vector fields in a
scalar-vector-tensor theory of gravity also leads to its
applications in modern cosmology and it might explain inflationary
scenarios
\cite{Lim:2004js,Kanno:2006ty,Watanabe:2009ct,Avelino:2009wj} and
accelerated expansion of the universe
\cite{Zlosnik:2006zu,Tartaglia:2007mh}. Based on a dynamical
vector field coupled to the gravitation and scalar fields, we have
studied to some extent the cosmological implications of a
scalar-vector-tensor theory of gravity \cite{:2007xt}. The models
also allow crossing of phantom divide line~\cite{Nozari:2008ff}.

Motivated by string theory and its extension M-theory, the
standard model particles may be confined on a hypersurface, called
brane, embedded in a higher dimensional space, called bulk. Only
gravity and other exotic matter such as the dilaton can propagate
in the bulk~\cite{Horava:1995qa}. The braneworld models have been
shown to be extremely rich in phenomena leading to modifications
of General Relativity (GR) at both low and high
energies~\cite{Maartens:2003tw}. In the context of gravity and
cosmology, models proposed by Randall and Sundrum (RS)
\cite{Randall:1999ee,Randall:1999vf} have attracted much
attention, where four-dimensional gravity can be recovered at low
energy despite the infinite size of the extra dimension. In RS II
model \cite{Randall:1999vf}, a positive tension brane is embedded
in five-dimensional anti-de Sitter (AdS) spacetime. To study
gravity on the brane, it is useful to derive the effective
four-dimensional Einstein equation on the brane firstly developed
by Shiromizu, Maeda, and Sasaki (SMS)~\cite{Shiromizu:1999wj}.
There are two very important results that arise from the effective
four-dimensional Einstein equations on the brane. The first one is
quadratic energy-momentum tensor, $\pi_{\mu\nu}$, which is
relevant in high energy and the second one is the projected Weyl
tensor, $E_{\mu\nu}$, on the brane which is responsible for
carrying on the brane the contribution of the bulk gravitational
field. In the RS II models, this term supplies an additional
matter-like effect to the brane. Thus, its contribution to the
four-dimensional effective theory is of crucial importance as it
is non-negligible already even in low energy limit. Then, the
Friedmann equations on the brane, governing the cosmological
evolution of the brane, are non conventional in that the Hubble
parameter depends quadratically on the energy density instead of
linearly as in standard cosmology, and one radiation like term,
usually referred to as a dark radiation term in the homogeneous
and isotropic background spacetime. This dark radiation modifies
the expansion of the background universe in the same way as an
usual radiation
\cite{Ida:1999ui,Kraus:1999it,Mukohyama:1999qx,Ichiki:2002eh}.

Recently, a braneworld scenario with bulk broken Lorentz
invariance has been developed, where a family of static
self-tuning braneworld solutions was found~\cite{Koroteev:2009xd}.
In a different approach braneworld model, a bulk vector field with
a non-vanishing vacuum expectation value, allowing for the
spontaneous breaking of the Lorentz symmetry. The breaking of
Lorentz invariance the loss of this symmetry is transmitted to the
gravitational sector of the model. By assuming that the vacuum
expectation value of the component of the vector field normal to
the brane vanishes, it found that Lorentz invariance on the brane
can be made exact via the dynamics of the graviton, vector field,
and the geometry of the extrinsic curvature of the surface of the
brane. As a consequence of the exact reproduction of Lorentz
symmetry on the brane, a condition for the matching of the
observed cosmological constant in four dimensions is
found~\cite{Bertolami:2006bf}. The notion of Lorentz violation in
four dimensions is extended to a five-dimensional braneworld
scenario resulting the time variation in the gravitational
coupling and cosmological constant. There exist also a relation
between the maximal velocity in the bulk and the speed of light on
the brane~\cite{Ahmadi:2006cr}. Various Lorentz violating effects
within the context of the braneworld scenario have also been
studied in
Refs.~\cite{Csaki:2000dm,Stoica:2001qe,Libanov:2005yf,Nozari:2008rg,Farakos:2009ui}.

In this paper we address the issue of cosmological evolution on a
brane in a theory of gravity whose action includes, in addition to
the familiar Einstein term, a Lorentz violating vector field
contribution. We generalize the gravitational effects of the
vector fields in four
dimensions~\cite{Jacobson:2000xp,Kostelecky:2003fs} to include
five dimensional braneworld gravity. In particular, we put a
vector $n^a$ in the direction of the extra dimension such that the
existence of the brane defines a preferred direction in the bulk.

This paper is organized as follows. In Section~\ref{sec:SMS
Effective Equation}, we derive the four-dimensional effective
Einstein equations on the brane in the case with the Lorentz
invariant violation in a bulk spacetime. With non-ignoring of the
Lorentz violation effects, this equation is modified by the trace
of the brane energy--momentum tensor. Thus the relation between
the projected Weyl tensor and the brane matter may be understood.
In Section~\ref{sec:Braneworld cosmology}, we study the
cosmological implications of the modified four-dimensional
effective Einstein equations on the brane. In general, the
effective four-dimensional Einstein equations on the brane cannot
be solved without knowing $E_{\mu\nu}$, because it could have a
non-trivial component of an anisotropic
stress~\cite{Maartens:2000fg}. However, it is possible to know
some features of this tensor by using constraint equations on the
brane obtained by the four-dimensional Bianchi identity. In the
background spacetime, the four-dimensional equations are
sufficient to show that $E_{\mu\nu}$ induces the radiation fluid
on the brane. We will take this strategy to determine the
Friedmann equation on the brane. Interestingly, the Friedmann
equation is found to depend on the equation of state of the matter
explicitly, and the Lorentz violation parameters. In Section
\ref{sec:LowEnergy}, we discuss a low energy limit of the theory.
Remarkable, the parameters of the theory can be determined by
equation of state of the brane matter. Section \ref{sec:discuss}
is devoted to the conclusions.

\section{\label{sec:SMS Effective Equation}Modified SMS Effective Equation on the brane}
In this section, we derive the $4$-dimensional effective
gravitational equations in a $Z_2$-symmetric braneworld using the
geometrical projection approach. For this purpose, we first write
the $5$-dimensional field equations in the form of the evolution
equations along the extra dimension and the constraint equations.

The action we consider consists of the vector field $n^{a}$
minimally coupled to gravity:
\begin{eqnarray}
    S &=&  \frac{1}{2\kappa^2}\int d^{5} x \sqrt{-\tilde{g}} \left({\cal R}
    -2\Lambda\right)+\int d^{5} x \sqrt{-\tilde{g}}{\cal L}_{n} +  \int d^{4}x
        \sqrt{-g}(-\sigma + {\cal L}_{m}) \ .
        \label{eq:action}
\end{eqnarray}
Here, $\mathcal{R}$, $\kappa$,  $\Lambda$, and $\tilde{g}$ are the
scalar curvature, the gravitational constant in $5$-dimensions,
the bulk cosmological constant, and the determinant of
$5$-dimensional metric, respectively. ${\cal L}_{m}$ and ${\cal
L}_{n}$ are the Lagrangian density for the matter fields on the
brane and the vector field Lagrangian, respectively. A metric $g$
is the induced metric on the brane while $\sigma$ denotes the
brane tension. Note that we have assumed no coupling between the
matter fields and the vector field in the action
(\ref{eq:action}). Therefore, the brane observer does not feel the
present of the preferred frame.

We write the coordinate system for the bulk spacetime in the form
\begin{equation}
    ds^2= g_{ab} dx^a dx^b = dy^2 +  g_{\mu\nu}(y,x)dx^\mu dx^\nu  \ ,
    \label{eq:metric}
\end{equation}
and we may assume that the position of the brane is $y=0$ in this
coordinate system so that the induced metric on the brane is
$g_{\mu\nu}(x)=\tilde{g}_{\mu\nu}(y=0,x)$. We also assume a
$Z_2$-symmetry across the brane and the extrinsic curvature is
defined as $K_{\mu\nu}=-g_{\mu\nu,y}/2$.

The vector field Lagrangian, ${\cal L}_{n}$, is given by
\begin{eqnarray}
    {\cal L}_{n} &=&-\beta_{1} \nabla^{a}n^{b}\nabla_{a}n_{b}- \beta_{2}\left(\nabla_{a}n^{a}\right)^2
    - \beta_{3}\nabla^{a}n^{b}\nabla_{b}n_{a}+\lambda (n^{a}n_{a}-1)  \ ,
    \label{eq:lagvector}
\end{eqnarray}
where $\beta_i$ are constant parameters and $\lambda$ is a
Lagrangian multiplier. In this setup, we assume that $n^a$ is a
vector field along the extra dimension and the preferred frame is
selected by the constrained vector field $n^{a}$ which violates
Lorentz symmetry. We take $n^a$ as the dimensionless vector.
Hence, each $\beta_i$ has dimension of $(mass)^3$. In other words,
$\beta_i^{1/3}$ gives the mass scale of symmetry breakdown in the
bulk. Following the usual braneworld scenarios our spacetime is
orthogonal to the extra dimension. Then one can introduce the
normal unit vector $n^{a}$ which is orthogonal to the
hypersurfaces at $y=const$. as $n^{a}=\delta_y^a$. In particular,
there is a background solution that 5-vector takes on a vacuum
expectation value with components  $(0,0,0,0,1)$, thus allowing
for the spontaneous breaking of the Lorentz symmetry.

Varying the action (\ref{eq:action}) with respect to the metric,
$\lambda$, and $n^a$, respectively, we have the field equations
\begin{eqnarray}
   {}^{(5)}G_{ab} &=& -\Lambda g_{ab}+ \kappa^2 (T_{ab}+{\cal T}_{ab})+ \kappa^2 \delta_a^\mu \delta_b^\nu S_{\mu\nu} \delta(y)  \ ,
   \label{eq:einstein-eq}\\
   g_{ab}n^an^b &=& 1 \ ,
     \label{eq:consvector}\\
   \nabla_{a}J^{ab} &=& \lambda n^b \ ,
     \label{eq:eosvector}
\end{eqnarray}
where current tensor $J^a{}_c$ is given by
\begin{equation}
    J^a{}_b =-\beta_{1} \nabla^{a}n_{b}- \beta_{2} \delta^a_b \nabla_{c}n^{c} - \beta_{3}\nabla_{b}n^{a} \ ,
    \label{eq:curenttensor}
\end{equation}
and $S_{\mu\nu}=-\sigma g_{\mu\nu}+\tau_{\mu\nu}$ is the energy
momentum tensor on the brane, where $\tau_{\mu\nu}$ is the energy
momentum tensor of the brane matter other than the tension.
$T_{ab}$ is the energy--momentum tensor of the vector field. To be
as general as possible, we also have included a bulk
energy--momentum tensor in (\ref{eq:einstein-eq}), denoted by
${\cal T}_{ab}$.

Using the extrinsic curvature, the components of the left hand
side of Einstein equations (\ref{eq:einstein-eq}) are
\begin{eqnarray}
   {}^{(5)}G^y{}_{y} &=& -{1\over 2}R + {1\over 2}K^2 - {1\over 2}K^{\alpha\beta}K_{\alpha\beta} =-\Lambda +\kappa^2 T^y{}_{y}+\kappa^2 {\cal T}^y{}_{y}  \ ,
   \label{eq:einstein-eq-yycomp}\\
   {}^{(5)}G^y{}_{\mu} &=& - D_\alpha K_{\mu}{}^{\alpha}+D_{\mu}K =\kappa^2 (T^y{}_{\mu}+{\cal T}^y{}_{\mu}) \ ,
   \label{eq:einstein-eq-ymucomp}\\
   {}^{(5)}G^\mu{}_{\nu} &=& G^\mu{}_{\nu} +
   (K^\mu{}_{\nu}-\delta^\mu_{\nu}K)_{,y}+ {1\over 2}\delta^\mu_{\nu}(K^2 + K^{\alpha\beta}K_{\alpha\beta})\nonumber\\
   &=&-\Lambda \delta^\mu_{\nu} +\kappa^2 (T^\mu{}_{\nu}+ {\cal T}^\mu{}_{\nu}) + \kappa^2 S^\mu{}_{\nu}\delta(y) \ ,
   \label{eq:einstein-eq-munucomp}
\end{eqnarray}
where $G^\mu{}_{\nu}$ is the $4$-dimensional Einstein tensor and
the covariant derivatives $D_\mu$ is calculated with respect to
the four-dimensional metric $g_{\mu\nu}$. The components of the
energy momentum tensor of the vector field are given by
\begin{eqnarray}
   T^y{}_{y} &=& \beta_2 K^2 + (\beta_1+\beta_3)K^{\alpha\beta}K_{\alpha\beta}  \ ,
   \label{eq:vector-eq-yycomp}\\
   T^y{}_{\mu} &=& 0\ ,
   \label{eq:vector-eq-ymucomp}\\
   T^\mu{}_{\nu} &=& 2(\beta_1+\beta_3)K^\mu{}_{\nu}K + \beta_2 \delta^\mu{}_{\nu}
   K^2 -\delta^\mu_\nu(\beta_1+\beta_3)K^{\alpha\beta}K_{\alpha\beta}
   - 2(\beta_1+\beta_3)K^\mu{}_{\nu,y} - 2\beta_2 \delta^\mu_\nu K_{,y} \ .
   \label{eq:vector-eq-munucomp}
\end{eqnarray}
Combining Eqs.~(\ref{eq:einstein-eq-yycomp}) with
(\ref{eq:einstein-eq-munucomp}) and using
(\ref{eq:vector-eq-yycomp}) and (\ref{eq:vector-eq-munucomp}), we
have
\begin{eqnarray}
    -{1\over 3}\left(R^\mu{}_{\nu} - {1\over 4}\delta^\mu_\nu R\right)
    &=& {1\over 6} \delta^\mu_\nu \Lambda  +{ (1-\alpha_0)\over 12}\delta^\mu_\nu K^2
    -{(1+\alpha_1)\over 3}\left(KK^\mu{}_{\nu}-{ 3 \over 4}\delta^\mu_\nu
    K_{\alpha\beta}K^{\alpha\beta}\right)\nonumber\\
  && +{(1+\alpha_1)\over 3}K^\mu{}_{\nu,y} - {(1-\alpha_0)\over 3}\delta^\mu_\nu K_{,y}
   -{\kappa^2 \over 3}\left({\cal T}^\mu{}_{\nu}-{1\over 2}\delta^\mu_\nu{\cal T}^y{}_{y}\right)\ ,
  \label{eq:tracelesEQ}
\end{eqnarray}
where we have defined
\begin{eqnarray}
   \alpha_0 = 2\kappa^2 \beta_2,\quad \alpha_1 = 2\kappa^2 (\beta_1+\beta_3)\ .
   \label{eq:pardefine}
\end{eqnarray}
The trace of equation (\ref{eq:tracelesEQ}) yields
\begin{eqnarray}
    &&(3-4\alpha_0-\alpha_1)K_{,y}=2\Lambda-(\alpha_0+\alpha_1)K^2
    +3(1+\alpha_1)K_{\alpha\beta}K^{\alpha\beta}-{\kappa^2 \over 3}\left({\cal T}^\mu{}_{\mu}-2{\cal T}^y{}_{y}\right) .
   \label{eq:traceK}
\end{eqnarray}
Substituting Eqs.~(\ref{eq:tracelesEQ}) and (\ref{eq:traceK}) into
the following components of the Weyl tensor
\begin{eqnarray}
   C_{y\mu y\nu}&=&-{1\over 3}\left(R_{\mu\nu}-{1\over 4} g_{\mu\nu}R\right)+{1\over 3}\left(K K_{\mu\nu}-{1\over 4} g_{\mu\nu}K^2\right)
   +{1 \over 3}\left(K_{\mu}{}^{\alpha}K_{\alpha\nu}+{3\over 4}g_{\mu\nu}K_{\alpha\beta}K^{\alpha\beta}\right)\nonumber\\
  &&+{2 \over 3}\left(K_{\mu\nu,y}- {1 \over 4} g_{\mu\nu}K_{,y}\right) \ ,
   \label{eq:weyltensor}
\end{eqnarray}
we have
\begin{eqnarray}
   \frac{3(1+\alpha_1)}{(3+\alpha_1)}C_{y\mu y\nu}&=& {1\over 2}\Lambda g_{\mu\nu}-\frac{3\alpha_0
   + (2+\alpha_0)\alpha_1}{4(3+\alpha_1)}g_{\mu\nu}K^2
   -\frac{(1+\alpha_1)\alpha_1}{(3+\alpha_1)}K K_{\mu\nu}
   +\frac{(1+\alpha_1)(3+2\alpha_1)}{(3+\alpha_1)}K_{\mu}{}^{\lambda}K_{\lambda\nu}\nonumber\\
  &&+\frac{3(1+\alpha_1)(4+\alpha_1)}{4(3+\alpha_1)}g_{\mu\nu}K_{\alpha\beta}K^{\alpha\beta}+(1+\alpha_1)K_{\mu\nu,y}
  - (1-\alpha_0)g_{\mu\nu} K_{,y}\nonumber\\
  &&  +{\kappa^2\over 2} g_{\mu\nu} {\cal T}^y{}_{y}-{\kappa^2\over 3} \left({\cal T}_{\mu\nu}
  + {1\over 2}g_{\mu\nu}{\cal T}^{\alpha}{}_{\alpha} \right) \ .
   \label{eq:weyltensor-1}
\end{eqnarray}
Here, we have defined that the term ${\cal T}^{\alpha}{}_{\alpha}$
is the trace defined with respect to the four-dimensional metric
$g$, and not the full trace defined with respect to $\tilde{g}$.
Equation~(\ref{eq:einstein-eq-munucomp}) can be expressed as
\begin{eqnarray}
   G_{\mu\nu}&=&-\Lambda g_{\mu\nu}+(1+\alpha_1)KK_{\mu\nu}-{(1-\alpha_0)\over 2}g_{\mu\nu}K^2
   -2(1+\alpha_1)K_{\mu}{}^{\alpha}K_{\alpha\nu}-{(1+\alpha_1)\over 2}g_{\mu\nu}K_{\alpha\beta}K^{\alpha\beta}\nonumber\\
  &&-(1+\alpha_1)K_{\mu\nu,y} + (1-\alpha_0)g_{\mu\nu} K_{,y}+\kappa^2 {\cal T}_{\mu\nu}\ .
  \label{eq:einstein-eq-1}
\end{eqnarray}
Using Eq.~(\ref{eq:weyltensor-1}), Eq.~(\ref{eq:einstein-eq-1}) is
expressed as
\begin{eqnarray}
   G_{\mu\nu}&=&-\frac{1}{2}\Lambda g_{\mu\nu}-\frac{3(1+\alpha_1)}{(3+\alpha_1)}E_{\mu\nu}
   -\frac{3(1+\alpha_1)}{(3+\alpha_1)}(K_{\mu}{}^{\alpha}K_{\alpha\nu}-KK_{\mu\nu})
   -\frac{6+4\alpha_1-(3+\alpha_1)\alpha_0}{4(3+\alpha_1)}g_{\mu\nu}K^2\nonumber\\
  &&+\frac{(1+\alpha_1)(6+\alpha_1)}{4(3+\alpha_1)}g_{\mu\nu}K_{\alpha\beta}K^{\alpha\beta}
  +{\kappa^2\over 2} g_{\mu\nu} {\cal T}^y{}_{y}+{2\kappa^2\over 3} \left({\cal T}_{\mu\nu}
  - {1\over 4}g_{\mu\nu}{\cal T}^{\alpha}{}_{\alpha} \right) \ ,
  \label{eq:einstein-eq-2}
\end{eqnarray}
where the projected Weyl tensor is $E_{\mu\nu}=C_{y\mu
y\nu}|_{y=0}$. Note that the coefficient of the four-dimensional
Einstein tensor (\ref{eq:einstein-eq-2}) is modified by factor
$(3+\alpha_1)$. Here, we take $\alpha_1 \neq -3$. The case
$\alpha_1 = -3$ provides a relation between the extrinsic
curvature and the projected Weyl tensor. To eliminate the
extrinsic curvature, we use the junction conditions. It can be
obtained by collecting together the terms in field equations which
contain a $\delta$-function. From
Eqs.~(\ref{eq:einstein-eq-munucomp}) and
(\ref{eq:vector-eq-munucomp}), we then obtain
\begin{eqnarray}
\left[ K^\mu{}_{\nu} - \delta^\mu_\nu K \right] |_{y=0}
    &=& {\kappa^2 \over 2(1+\alpha_1)}  \left( S^\mu{}_{\nu}   + \alpha_2 \delta^\mu_\nu S \right) \ ,
    \label{eq:JC}
\end{eqnarray}
where
\begin{eqnarray}
   \alpha_2 =  \frac{\alpha_0+\alpha_1}{3-4\alpha_0-\alpha_1}\ .
\end{eqnarray}
For convenient we will take $\alpha_1 \neq -3$ and $\alpha_1 \neq
-1$ in order to avoid unreal singularities in
Eqs.~(\ref{eq:einstein-eq-2}) and (\ref{eq:JC}). Substituting
(\ref{eq:JC}) into (\ref{eq:einstein-eq-2}), we finally obtain the
modified effective SMS equation on the brane as
\begin{eqnarray}
   G_{\mu\nu} &=& -\Lambda_b g_{\mu\nu}+8\pi G \left(\tau_{\mu\nu}+{\alpha_1\over 12} g_{\mu\nu}\tau\right)
   + \kappa^4 \pi_{\mu\nu}- \widetilde{E}_{\mu\nu}+ F_{\mu\nu}  \ ,
   \label{eq:einstein-eq-brane}
\end{eqnarray}
where we have defined the quantities
\begin{eqnarray}
   \Lambda_b&=&\frac{1}{2}\Lambda+\frac{\kappa^4}{4(3-4\alpha_0-\alpha_1)}\sigma^2,
   \label{eq:lambdabrane}\\
   8\pi G&=&\frac{3\kappa^4}{2(3+\alpha_1)(3-4\alpha_0-\alpha_1)}\sigma,
    \label{eq:effgravconstan}\\
   \pi_{\mu\nu}&=&\frac{3}{4(3+\alpha_1)(1+\alpha_1)}\left[{(1-2\alpha_0-\alpha_1)\over (3-4\alpha_0-\alpha_1)}\tau\tau_{\mu\nu}- \tau_{\mu}{}^{\alpha}\tau_{\alpha\nu} +{(6+\alpha_1)\over 12}g_{\mu\nu}\tau_{\alpha\beta}\tau^{\alpha\beta}\right.  \nonumber\\
   &&\left. - {2(3-\alpha_1) -(9+\alpha_1)\alpha_0 \over
   12(3-4\alpha_0-\alpha_1)}g_{\mu\nu}\tau^2\right]\ ,
    \label{eq:quadraticem}\\
   \widetilde{E}_{\mu\nu}&=& \frac{3(1+\alpha_1)}{(3+\alpha_1)}E_{\mu\nu} \ ,
   \label{eq:redefineweyltensor}
\end{eqnarray}
and the bulk energy-momentum tensor projected on the brane is
given by
\begin{eqnarray}
   F_{\mu\nu}&=& \left[{\kappa^2\over 2} g_{\mu\nu} {\cal T}^y{}_{y} +{2\kappa^2\over 3} \left({\cal T}_{\mu\nu}
  - {1\over 4}g_{\mu\nu}{\cal T}^{\alpha}{}_{\alpha} \right)\right]_{y=0} \ .
  \label{eq:bulkfield}
\end{eqnarray}
There are four features in the effective Einstein equations
(\ref{eq:einstein-eq-brane}). The first one is the presence of the
bulk energy-momentum tensor. This term allows exotic matter such
as the dilaton can propagate in the bulk. The second departure
from the standard four-dimensional Einstein equation arises from
the presence of the Weyl tensor which is undetermined on the
brane. The third is a quadratic in the brane energy-momentum
tensor. The last one is a linear in addition to the brane
energy-momentum tensor. It is our main result. This trace part of
the brane energy-momentum tensor is measured by local observers at
the brane and vanishes when
$\alpha_1=2\kappa^2(\beta_1+\beta_3)=0$.

Equation (\ref{eq:einstein-eq-ymucomp}) and the junction
conditions (\ref{eq:JC}) imply
\begin{eqnarray}
     D_\mu \tau^{\mu}{}_{\nu} + \alpha_2 D_\nu \tau -(1+4\alpha_2)D_\nu \sigma=-2(1+\alpha_1) {\cal T}^y{}_{\nu}\ .
    \label{eq:continuity}
\end{eqnarray}
This equation tell us that the energy momentum tensor
$\tau_{\mu\nu}$ is not conserved on the brane. Taking the
divergence of the four--dimensional effective equations and using
four--dimensional Bianchi identity, we obtain the constraint
equations for $E_{\mu\nu}$ as
\begin{eqnarray}
    D_\mu\widetilde{E}^{\mu}{}_{\nu} &=&-D_\nu\Lambda_b +8\pi G \left(D_\mu\tau^{\mu}{}_{\nu} +{\alpha_1\over 12}D_\nu \tau \right)
    + \kappa^4 D_\mu \pi^{\mu}{}_{\nu} \nonumber\\
    &&+{\kappa^2\over 2} D_\nu {\cal T}^y{}_{y}+{2\kappa^2\over 3} \left(D_\mu{\cal T}_{\mu\nu}
  - {1\over 4}D_\nu{\cal T}^{\alpha}{}_{\alpha} \right).
    \label{eq:gencontinuityWeyl}
\end{eqnarray}
Equations (\ref{eq:continuity}) and (\ref{eq:gencontinuityWeyl})
indicate a time variation of the brane tension, the cosmological
constant, and the gravitational constant in general.

In the following section, we study analytically the cosmological
consequences of Eqs.~(\ref{eq:einstein-eq-brane}),
(\ref{eq:continuity}) and (\ref{eq:gencontinuityWeyl}). Here, for
simplicity, we consider constant $\sigma$, because there are no
theoretical observational arguments for the evolution of $\sigma$
in time. For cosmology on the brane, we suppose here that we can
ignore the bulk matter, $F_{\mu\nu}=0$. The bulk matter is
important to get a well--behaved geometry in the bulk. We also
assume that the bulk cosmological constant is truly constant.
Then, Eqs.~(\ref{eq:continuity}) and (\ref{eq:gencontinuityWeyl})
become
\begin{eqnarray}
    &&D_\mu\tau^{\mu}{}_{\nu} =- \alpha_2 D_\nu \tau  \ ,
    \label{eq:continuity1}\\
    &&D_\mu\widetilde{E}^{\mu}{}_{\nu} = -8\pi G \left(\alpha_2 -{\alpha_1\over 12} \right)D_\nu \tau
    + \kappa^4 D_\mu \pi^{\mu}{}_{\nu}\ .
    \label{eq:continuityWeyl}
\end{eqnarray}
Note that the projected Weyl tensor is affected by the
energy--momentum tensor on the brane even at low energies. Thus,
the model is quite different from the conventional braneworld even
at low energies.

\section{\label{sec:Braneworld cosmology}Braneworld cosmology}
The projected Weyl tensor in the modified Einstein equation
(\ref{eq:einstein-eq-brane}) is a priori undetermined. This comes
from the five-dimensional nature of the theory and the fact that
the system of equations is not closed on the brane. This tensor
mediates some information from the bulk to the brane. In this
section, we will try to solve Einstein equation to study the
cosmology braneworld from equation (\ref{eq:einstein-eq-brane}),
by assuming that there is no cosmological constant on the brane
and the constant vacuum energy. Although these assumptions are
usual in braneworld scenario, we will show, which is the main
result of present paper, the effective Friedmann equations is
modified by the effect of Lorentz violation, and the components of
the projected Weyl tensor are related to the matter on the brane.
We then discuss the method to obtain the components of the
projected Weyl tensor from the brane data. For cosmological
applications, we consider a metric of the form
\begin{eqnarray}
    ds^2  = -dt^2 + a^2(t) \delta_{ij} dx^i dx^j \ ,
    \label{eq:metriccos}
\end{eqnarray}
where $x^i$ are the three ordinary spatial coordinates and $a$ is
the scale factor. The Hubble parameter $H$ on the brane,
describing the cosmological dynamics of the Universe, is defined
as $H = \dot{a}/a$. For simplicity, we ignore the bulk matter for
the cosmology on the brane. Hereafter, we will consider only the
matter on the brane. For further discussions on the gravitational
field equations in the braneworld model with Lorentz violation and
their cosmological applications see \cite{Ahmadi:2006cr}. We
restrict the energy-momentum tensor on the brane of the form
\begin{eqnarray}
    \tau_{\mu\nu} = (\rho, P a^2 \delta_{ij}) \ ,
    \label{eq:EMbrane}
\end{eqnarray}
where $\rho$ is the energy density and $P$ the pressure. We will
assume that the equation of state relating $\rho$ and $P$ has the
form $P=\omega\rho$, where $\omega$ is constant. Similarly, the
projected Weyl tensor is of the form
\begin{eqnarray}
    E_{\mu\nu} = (\rho_d, P_d a^2 \delta_{ij}) \ .
    \label{eq:weyltensorcos}
\end{eqnarray}
The traceless property of $E_{\mu\nu}$ implies: $-\rho_d+3P_d=0$.
We will be interested in the relation between the components of
the projected Weyl tensor and the brane energy-momentum tensor.
The components of the quadratic in the energy-momentum tensor
(\ref{eq:quadraticem}) are given by
\begin{eqnarray}
    \pi_{00}&=&\frac{1+3\alpha_3}{4(3+\alpha_1)(1+\alpha_1)^2}\rho^2 \ ,
    \label{eq:quadratic00}\\
    \pi_{ij}&=&\frac{1+2\omega-3\alpha_4}{4(3+\alpha_1)(1+\alpha_1)^2}\rho^2 a^2\delta_{ij} \ ,
    \label{eq:quadraticij}
\end{eqnarray}
where
\begin{eqnarray}
   \alpha_3 &=& \frac{1}{12(3-4\alpha_0-\alpha_1)}\{[7-9(2+\omega)\omega
   -3(1+\omega)^2 (2-\alpha_1)\alpha_1]\alpha_0 - [17-3\omega(8+3\omega)\nonumber\\
   &&+2(1-12\omega-3\omega^2+(1+3\omega^2)\alpha_1^2)]\alpha_1\}\ ,\nonumber\\
   \alpha_4 &=&\frac{1}{12(3-4\alpha_0-\alpha_1)}\{ [-1-2\omega+15\omega^2
   +3(1+\omega)^2(6+\alpha_1)\alpha_1]\alpha_0-[15+32\omega\nonumber\\
   &&-15\omega^2-(1+3\omega^2)\alpha_1^2 -
   2(1+9\omega^2)\alpha_1]\alpha_1\} \ .
\end{eqnarray}
Substituting metric (\ref{eq:metriccos}) and tensors
(\ref{eq:EMbrane}), (\ref{eq:weyltensorcos}) and
(\ref{eq:quadratic00}), (\ref{eq:quadraticij}) in the effective
Einstein equations (\ref{eq:einstein-eq-brane}), one finds
\begin{eqnarray}
    &&3H^2=8\pi G \left[1+\frac{(1-3\omega)\alpha_1}{12} \right]\rho
    +\frac{\kappa^4(1+3\alpha_3)}{4(1+\alpha_1)^2(3+\alpha_1)}\rho^2 - \frac{3(1+\alpha_1)}{(3+\alpha_1)}\rho_d \ ,
    \label{eq:friedmann}\\
    &&-2\dot{H}-3H^2=8\pi G \left[\omega-\frac{(1-3\omega)\alpha_1)}{12} \right]\rho
    +\frac{\kappa^4(1+2\omega-3\alpha_4)}{4(1+\alpha_1)^2(3+\alpha_1)}\rho^2 - \frac{(1+\alpha_1)}{(3+\alpha_1)}\rho_d\ .
    \label{eq:raychauduri}
\end{eqnarray}
Obviously, these equations are quite different from the usual
braneworld equations due to the effect of bulk Lorentz violation.
From Eq.~(\ref{eq:continuity1}) and the constraint equation for
the projected Weyl tensor (\ref{eq:continuityWeyl}), we have
\begin{eqnarray}
    &&[1+(1-\omega)\alpha_2]\dot{\rho} + 3H\rho(1+\omega)=0 \ ,
    \label{eq:conservationcos}
\end{eqnarray}
and
\begin{eqnarray}
   \dot{\rho}_d + 4H\rho_d&=&\frac{8\pi
    G(1+\omega)(1-3\omega)(3+\alpha_1)(3-4\alpha_0-\alpha_1)}{3(1+\alpha_1)^3}
    \left[\alpha_2 -\frac{(1+\alpha_1)^2\alpha_1}{12(1-\omega\alpha_1-(1+\omega)\alpha_0)} \right]H\rho \nonumber\\
    &&-\frac{\kappa^4(1+\omega)\alpha_{5}}{4(1+\alpha_1)^3(1+(1-3\omega)\alpha_2)}H\rho^2,
    \label{eq:conservationcosweyl}
\end{eqnarray}
where
\begin{eqnarray}
   \alpha_5&=&\frac{(1+\alpha_1)}{2(3-4\alpha_0-\alpha_1)}\{12(1+\omega)^2\alpha_0^2\alpha_1
   +[9(1+3\omega^2)+(1+3\omega^2)\alpha_1^2-2(1+12\omega-9\omega^2)\alpha_1]\alpha_1\nonumber\\
   &&+[3(1-3\omega^2)-2(7+30\omega-9\omega^2)\alpha_1+(7+6\omega+15\omega^2)\alpha_1^2]\alpha_0\} \ .
\end{eqnarray}
For $\omega\neq -1$, equation (\ref{eq:conservationcos}) is solved
to yield
\begin{eqnarray}
    \rho = a^{-\frac{3(1+\omega)}{1+(1-\omega)\alpha_2}} \ .
\end{eqnarray}
Here, we have absorbed a constant factor into the scale factor by
rescaling it. Equation (\ref{eq:conservationcosweyl}) can be
integrated. We find
\begin{eqnarray}
    {\rho}_d &=& -{3C\over a^4}+\frac{8\pi G(1+\omega)(3+\alpha_1)(3-4\alpha_0-\alpha_1)^2}{9(1+\alpha_1)^4}
    \left\{ [1+(1-3\omega)\alpha_2]\alpha_2
    -\frac{\alpha_1(1+\alpha_1)^2}{4(3-4\alpha_0-\alpha_1)}\right\}a^{-\frac{3(1+\omega)}{1+(1-\omega)\alpha_2}}\nonumber\\
    &&-\frac{\kappa^4 (1+\omega)\alpha_{5}}{8(1+\alpha_1)^3[2(1-3\omega)\alpha_2-(1+3\omega)]}a^{-\frac{6(1+\omega)}{1+(1-\omega)\alpha_2}} \ ,
    \label{eq:compweyl-sol}
\end{eqnarray}
where $C$ is a constant of integration. This effect of the bulk
acts as radiation fluid, hence it is called as dark radiation.
Substituting Eq.~(\ref{eq:compweyl-sol}) into
Eq.~(\ref{eq:friedmann}), we obtain the effective Friedmann
equation
\begin{eqnarray}
    H^2&=&{8\pi G_{eff}\over 3}\rho+ A \rho^2 + {\bar{C}\over a^4} \ ,
    \label{eq:friedmann-special}
\end{eqnarray}
where
\begin{widetext}
\begin{eqnarray}
   G_{eff}&=&\left\{1 - \frac{[1-\omega\alpha_1-(1+\omega)\alpha_0][(2+3\omega-(2+\alpha_1)\alpha_1)\alpha_1+3(1+\omega)\alpha_0]}{3(1+\alpha_1)^3}\right\}G  \ ,
    \label{eq:effNewton}\\
   A&=&
   \frac{\kappa^4}{12(3+\alpha_1)(1+\alpha_1)^2}\left[1+3\alpha_3-\frac{3(1+\omega)\alpha_{5}}{2[(1+3\omega)-2(1-3\omega)\alpha_2]}\right],
    \label{eq:Adefine}\\
   \bar{C}&=&\frac{3(1+\alpha_1)}{(3+\alpha_1)} C \ .
   \label{eq:Cdefine}
\end{eqnarray}
\end{widetext}
Note that the effective Newton constant depends on the Lorentz
violating parameters and the equations of state. It is different
from the conventional braneworld cosmology in five-dimensional
case even at low energy. If the effects of Lorentz violations are
ignored, $\beta_i=0$, we have $G_{eff}=G$, $A =\kappa^4/36$ and
$\bar{C}=C$. In the alternative theory of gravity including
Brans--Dicke theory, the effective Newton constant need not be
constant in time. Observational bounds on $\dot{G}/G$ then
constrain the theory. In our case, we have the relation
(\ref{eq:effNewton}), hence the Newton constant is always
constant.

If the effective cosmological constant is included, the Friedmann
equation (\ref{eq:friedmann-special}) becomes
\begin{eqnarray}
    H^2&=&{1\over 3}\Lambda_b + {8\pi G_{eff}\over 3}\rho+ A \rho^2 + {\bar{C}\over a^4} \ ,
    \label{eq:friedmann-special-1}
\end{eqnarray}
where the relation between the vacuum energy and the effective
cosmological constant on a brane is given by
Eq.~(\ref{eq:lambdabrane}). It is different from the usual
four-dimensional theory. In the RS braneworld, the vacuum energy
in the brane is not directly related to the cosmological constant
on the brane in the effective Einstein equation as in
Eq.~(\ref{eq:lambdabrane}). In the RS braneworld, there should be
a cancellation between the four-dimensional and five-dimensional
contribution of the vacuum energy in order to have a vanishing
cosmological constant on the brane. This requires a fine-tuning
for the parameters in the action. In the present model the RS type
relation is given by
\begin{eqnarray}
    \sigma = \frac{6}{\kappa^2 l}\left(1-{4\over 3}\alpha_0-{1\over 3}\alpha_1\right)^{1/2} \ ,
    \label{eq:finetunning}
\end{eqnarray}
and
\begin{eqnarray}
    8\pi G=\frac{3\kappa^2}{l(3+\alpha_1)(1+\alpha_1)^{1/2}}  \ .
    \label{eq:effgravconstan-ft}
\end{eqnarray}
Here, the bulk cosmological constant is defined as $\Lambda=
-6/\kappa^2 l^2$, where $l$ is the scale of the bulk curvature
radius.

\section{\label{sec:LowEnergy}Low Energy Constraint on $\beta_i$}
For a well-defined theory, the constraints on the theory
parameters $\beta_i$ are given by \cite{Lim:2004js} (see also
\cite{Carroll:2004ai}):
\begin{enumerate}
    \item Subluminal propagation of spin-0 field: $(\beta_{1}+\beta_{2}+\beta_{3})/\beta_{1}\leq
    1$,
    \item Positivity of Hamiltonian: $\beta_{1}>0$,
    \item Non-tachyonic propagation of spin-0 field:
        $(\beta_{1}+\beta_{2}+\beta_{3})/\beta_{1}\geq 0$,
    \item Subluminal propagation of spin-2 field:
        $\beta_{1}+\beta_{3}\leq 0$.
\end{enumerate}
All these conditions together imply
$(\beta_{1}+\beta_{2}+\beta_{3})\geq 0$ and $\beta_2\geq 0$.

At low energies, we can neglect the quadratic term of the
Friedmann equation (\ref{eq:friedmann-special}). Then we have
\begin{eqnarray}
    H^2&=&{8\pi G_{eff}\over 3}\rho + {\bar{C}\over a^4} \ .
    \label{eq:friedmann-special-le}
\end{eqnarray}
Here, we have assumed $3A/8\pi G_{eff}<<1$. Therefore, one can set
$A \approx 0$ without loss of generality. Solving
Eq.~(\ref{eq:Adefine}) one finds
\begin{eqnarray}
    \alpha_1 = \frac{1-\alpha_0(1+\omega)}{\omega},
    ~~\text{or}~~
       \alpha_1 =  \frac{2(1+3\omega)-3\alpha_0(1+\omega)^2}{1+3\omega^2}\ ,
    \label{eq:lowconst-2}
\end{eqnarray}
where $\alpha_0$ and $\alpha_1$ is given by
Eq.~(\ref{eq:pardefine}). In other word, the effect of Lorentz
violation in the bulk is dependent on the equation of state of the
energy components of the Universe. Remarkable, the first solution
(\ref{eq:lowconst-2}) yields $G_{eff}=G$. In this case, using the
above constraints we find
\begin{enumerate}
    \item For $\omega<-1$,
    \begin{eqnarray}
    \alpha_0 > \frac{1+3\omega}{1+\omega}, \qquad \alpha_{1}<
    -3\ ,
    \end{eqnarray}
    and
    \begin{eqnarray}
    1 < \alpha_0< \frac{1+3\omega}{1+\omega}, \qquad -3 < \alpha_{1}< -1 \ .
    \end{eqnarray}
    \item For $-1<\omega<0$,
    \begin{eqnarray}
    1< \alpha_0 \leq \frac{1}{1+\omega}, \qquad -1 < \alpha_{1}\leq 0\
    .
    \end{eqnarray}
    \item For $\omega>0$,
    \begin{eqnarray}
    \frac{1}{1+\omega} \leq \alpha_0< 1, \qquad -1 < \alpha_{1}\leq 0 \ .
    \end{eqnarray}
\end{enumerate}
The above constraints give the correction in the coefficient of
the dark radiation. The second solution (\ref{eq:lowconst-2})
gives the constraints:
\begin{enumerate}
    \item For $\omega<-1$,
    \begin{eqnarray}
    \alpha_0 > \frac{5+6\omega+9\omega^2}{3(1+\omega)^2}, \qquad \alpha_{1}<
    -3\ ,
    \end{eqnarray}
    and
    \begin{eqnarray}
    1 < \alpha_0< \frac{5+6\omega+9\omega^2}{3(1+\omega)^2}, \qquad -3 < \alpha_{1}< -1 \ .
    \end{eqnarray}
    \item For $-1< \omega \leq -1/3$,
    \begin{eqnarray}
    \alpha_0 > \frac{5+6\omega+9\omega^2}{3(1+\omega)^2}, \qquad \alpha_{1}<
    -3\ ,
    \end{eqnarray}
    and
    \begin{eqnarray}
    1 < \alpha_0< \frac{5+6\omega+9\omega^2}{3(1+\omega)^2}, \qquad -3 < \alpha_{1}< -1 \ .
    \end{eqnarray}
    \item For $\omega \geq -1/3$,
    \begin{eqnarray}
    \frac{2(1+3\omega)}{3(1+\omega)^2} \leq \alpha_0< 1, \qquad -1 < \alpha_{1}\leq 0 \ .
    \end{eqnarray}
\end{enumerate}
These constraints give the corrections both in the effective
Newton constant and the dark radiation.

\section{\label{sec:discuss}Conclusions}
In the present paper, we have considered a five-dimensional
braneworld model with bulk Lorentz invariance violation, and
derived the effective four-dimensional Einstein equations on the
brane. The main result of this paper is the existence of the trace
part of the brane energy-momentum tensor in the modified Einstein
equations on the brane, which is a modification of the SMS
effective equation~\cite{Shiromizu:1999wj}. Thus, the divergence
of the projected Weyl tensor is modified. Therefore, due to
Lorentz violating effect, we have obtained an expression for the
projected Weyl tensor as a function of the source on the brane. It
becomes clear that the bulk effect can be determined by matter
localized on the brane even at low energies. As an application, we
have used the modified SMS effective equation to determine the
Friedmann equation on the brane. We have showed the effective
Newton constant that relates geometry to the matter density in
Friedmann equation is dependent on the equation of state of the
energy component of the Universe, and the Lorentz violating
parameters. Note that if the brane was isotropic and homogeneous,
the matter part would have the additional property,
$D^{\mu}\pi_{\mu\nu}=0$. However, due to effect of Lorentz
violation in the bulk, the effect of matter still appears in
Eq.~(\ref{eq:continuityWeyl}). Thus, the brane matter will deform
the bulk geometry. In other word, the back-reaction of this to the
brane will modify the effective Friedmann equation even at low
energies. It is interesting to understand the low energy
description of this braneworld model. The low energy perturbation
scheme proposed in~\cite{Kanno:2006ty} is a major achievement as
it allows for the derivation of the effective theory on the brane
and for the full comprehension of the Weyl tensor contribution to
the effective theory. We leave this issue for future studies.

Finally, we also find that the effect of Lorentz violation in the
bulk is dependent on the equation of state of the energy
components of the brane matter. This model also provides a
convenient framework within which one may study dark energy.

\begin{acknowledgements}
Arianto wishes to acknowledge all members of the Theoretical
Physics Laboratory, the THEPI Divison of the Faculty of
Mathematics and Natural Sciences, ITB, for the warmest
hospitality. This work was supported by Hibah Kompetensi DIKTI,
2009.
\end{acknowledgements}


\begin{thebibliography}{99}
\bibitem{Kostelecky:1988zi}
  V.~A.~Kostelecky and S.~Samuel,
  Phys.\ Rev.\ D {\bf 39}, 683 (1989).

\bibitem{Bekenstein:2004ne}
  J.~D.~Bekenstein,
  Phys.\ Rev.\  D {\bf 70}, 083509 (2004)
  [Erratum-ibid.\  D {\bf 71}, 069901 (2005)]
  [arXiv:astro-ph/0403694].

\bibitem{Skordis:2005xk}
  C.~Skordis, D.~F.~Mota, P.~G.~Ferreira and C.~Boehm,
  Phys.\ Rev.\ Lett.\  {\bf 96}, 011301 (2006)
  [arXiv:astro-ph/0505519]
  ;
  C.~Skordis,
  Phys.\ Rev.\  D {\bf 74}, 103513 (2006)
  [arXiv:astro-ph/0511591].


\bibitem{Jacobson:2000xp}
  T.~Jacobson and D.~Mattingly,
  Phys.\ Rev.\  D {\bf 64}, 024028 (2001)
  [arXiv:gr-qc/0007031].

\bibitem{Zlosnik:2006sb}
  T.~G.~Zlosnik, P.~G.~Ferreira and G.~D.~Starkman,
  Phys.\ Rev.\  D {\bf 74}, 044037 (2006)
  [arXiv:gr-qc/0606039].


\bibitem{Zlosnik:2006zu}
  T.~G.~Zlosnik, P.~G.~Ferreira and G.~D.~Starkman,
  Phys.\ Rev.\  D {\bf 75}, 044017 (2007)
  [arXiv:astro-ph/0607411].


\bibitem{Bonvin:2007ap}
  C.~Bonvin, R.~Durrer, P.~G.~Ferreira, G.~Starkman and T.~G.~Zlosnik,
  Phys.\ Rev.\  D {\bf 77}, 024037 (2008)
  [arXiv:0707.3519 [astro-ph]].

\bibitem{Carroll:2004ai}
  S.~M.~Carroll and E.~A.~Lim,
  Phys.\ Rev.\  D {\bf 70}, 123525 (2004)
  [arXiv:hep-th/0407149].

\bibitem{Li:2007vz}
  B.~Li, D.~Fonseca Mota and J.~D.~Barrow,
  Phys.\ Rev.\  D {\bf 77}, 024032 (2008)
  [arXiv:0709.4581 [astro-ph]].


\bibitem{Kostelecky:2003fs}
  V.~A.~Kostelecky,
  Phys.\ Rev.\  D {\bf 69}, 105009 (2004)
  [arXiv:hep-th/0312310].

\bibitem{Kanno:2006ty}
  S.~Kanno and J.~Soda,
  Phys.\ Rev.\  D {\bf 74}, 063505 (2006)
  [arXiv:hep-th/0604192].

\bibitem{Watanabe:2009ct}
  M.~a.~Watanabe, S.~Kanno and J.~Soda,
  arXiv:0902.2833 [hep-th].

\bibitem{Avelino:2009wj}
  P.~P.~Avelino, D.~Bazeia, L.~Losano, R.~Menezes and J.~J.~Rodrigues,
  arXiv:0903.5297 [astro-ph.CO].

\bibitem{Lim:2004js}
  E.~A.~Lim,
  Phys.\ Rev.\  D {\bf 71}, 063504 (2005)
  [arXiv:astro-ph/0407437].

\bibitem{Tartaglia:2007mh}
  A.~Tartaglia and N.~Radicella,
  Phys.\ Rev.\  D {\bf 76}, 083501 (2007)
  [arXiv:0708.0675 [gr-qc]].

\bibitem{:2007xt}
  Arianto, F.~P.~Zen, B.~E.~Gunara, Triyanta and Supardi,
  JHEP {\bf 0709}, 048 (2007)
  [arXiv:0709.3688 [hep-th]];
  Arianto, F.~P.~Zen, Triyanta and B.~E.~Gunara,
  Phys.\ Rev.\  D {\bf 77}, 123517 (2008)
  [arXiv:0801.0331 [hep-th]];
  F.~P.~Zen, Arianto, B.~E.~Gunara, Triyanta and A.~Purwanto,
  arXiv:0809.3847 [hep-th].



\bibitem{Nozari:2008ff}
  K.~Nozari and S.~D.~Sadatian,
  Eur.\ Phys.\ J.\  C {\bf 58}, 499 (2008)
  [arXiv:0809.4744 [gr-qc]].

\bibitem{Horava:1995qa}
  P.~Horava and E.~Witten,
  Nucl.\ Phys.\  B {\bf 460}, 506 (1996)
  [arXiv:hep-th/9510209].


\bibitem{Maartens:2003tw}
  For a review see, e.g., R.~Maartens,
  Living Rev.\ Rel.\  {\bf 7}, 7 (2004)
  [arXiv:gr-qc/0312059].

\bibitem{Randall:1999ee}
  L.~Randall and R.~Sundrum,
  Phys.\ Rev.\ Lett.\  {\bf 83}, 3370 (1999)
  [arXiv:hep-ph/9905221].

\bibitem{Randall:1999vf}
  L.~Randall and R.~Sundrum,
  Phys.\ Rev.\ Lett.\  {\bf 83}, 4690 (1999)
  [arXiv:hep-th/9906064].

\bibitem{Shiromizu:1999wj}
  T.~Shiromizu, K.~i.~Maeda and M.~Sasaki,
  Phys.\ Rev.\  D {\bf 62}, 024012 (2000)
  [arXiv:gr-qc/9910076].

\bibitem{Ida:1999ui}
  D.~Ida,
  JHEP {\bf 0009}, 014 (2000)
  [arXiv:gr-qc/9912002].


\bibitem{Kraus:1999it}
  P.~Kraus,
  JHEP {\bf 9912}, 011 (1999)
  [arXiv:hep-th/9910149].

\bibitem{Mukohyama:1999qx}
  S.~Mukohyama,
  Phys.\ Lett.\  B {\bf 473}, 241 (2000)
  [arXiv:hep-th/9911165].

\bibitem{Ichiki:2002eh}
  K.~Ichiki, M.~Yahiro, T.~Kajino, M.~Orito and G.~J.~Mathews,
  Phys.\ Rev.\  D {\bf 66}, 043521 (2002)
  [arXiv:astro-ph/0203272].

\bibitem{Koroteev:2009xd}
  P.~Koroteev and M.~Libanov,
  Phys.\ Rev.\  D {\bf 79}, 045023 (2009)
  [arXiv:0901.4347 [hep-th]];
  P.~Koroteev and M.~Libanov,
  JHEP {\bf 0802}, 104 (2008)
  [arXiv:0712.1136 [hep-th]].

\bibitem{Bertolami:2006bf}
  O.~Bertolami and C.~Carvalho,
  Phys.\ Rev.\  D {\bf 74}, 084020 (2006)
  [arXiv:gr-qc/0607043].

\bibitem{Ahmadi:2006cr}
  F.~Ahmadi, S.~Jalalzadeh and H.~R.~Sepangi,
  Class.\ Quant.\ Grav.\  {\bf 23}, 4069 (2006)
  [arXiv:gr-qc/0605038];
  F.~Ahmadi, S.~Jalalzadeh and H.~R.~Sepangi,
  Phys.\ Lett.\  B {\bf 647}, 486 (2007)
  [arXiv:gr-qc/0702103].

\bibitem{Csaki:2000dm}
  C.~Csaki, J.~Erlich and C.~Grojean,
  Nucl.\ Phys.\  B {\bf 604}, 312 (2001)
  [arXiv:hep-th/0012143].

\bibitem{Stoica:2001qe}
  H.~Stoica,
  JHEP {\bf 0207}, 060 (2002)
  [arXiv:hep-th/0112020].

\bibitem{Libanov:2005yf}
  M.~V.~Libanov and V.~A.~Rubakov,
  JCAP {\bf 0509}, 005 (2005)
  [arXiv:astro-ph/0504249].

\bibitem{Nozari:2008rg}
  K.~Nozari and S.~D.~Sadatian,
  JCAP {\bf 0901}, 005 (2009)
  [arXiv:0810.0765 [gr-qc]].

\bibitem{Farakos:2009ui}
  K.~Farakos,
  arXiv:0903.3356 [hep-th].

\bibitem{Maartens:2000fg}
  R.~Maartens,
  Phys.\ Rev.\  D {\bf 62}, 084023 (2000)
  [arXiv:hep-th/0004166].



\end{thebibliography}
\end{document}